\documentclass[12pt,preprint]{aastex}
\shorttitle{NGC7538}
\shortauthors{Sandell et al.}

\begin{document}

\def\arcmin{{$^{\prime}$}}
\def\arcsec{{$^{\prime\prime}$}}
\def\ptsec{$''\mskip-7.6mu.\,$}
\def\psec{$^s\mskip-7.6mu.\,$}
\def\Msun{\,{\rm M$_{\odot}$}}
\def\Lsun{\,{\rm L$_{\odot}$}}
\def\ltsim{$\stackrel{<}{\sim}$}
\def\gtsim{$\stackrel{>}{\sim}$}
\def\jtra#1#2{${\rm J}\!\!=\!\!{#1}\!\!\to\!\!{#2}$}

\def\degr{$^{\circ}$}

\title{NGC\,7538\,S - a High-Mass Protostar with a Massive Rotating Disk}

\author{G\"oran Sandell\altaffilmark{1}, Melvyn Wright\altaffilmark{2}, James R. Forster\altaffilmark{2}}

\altaffiltext{1}{Universities Space Research Association, NASA Ames Research Center, MS 144-2,
Moffett Field, CA 94035, U.S.A.}
\email{gsandell@mail.arc.nasa.gov}
\altaffiltext{2}{Radio Astronomy Laboratory, University of California, Berkeley
601 Campbell Hall, Berkeley, CA 94720, U.S.A}

\begin{abstract}

We report the detection of a massive rotating disk around the high-mass
Class 0 candidate NGC\,7538\,S.  The disk is well-resolved with BIMA
($\theta_A$ =3\ptsec7) in 3.4~mm continuum and in H$^{13}$CN \jtra10. It
is seen nearly edge on and has a size of $\sim$30,000 AU.  A
young, powerful outflow perpendicular to the rotating disk is mapped in SiO
\jtra21 and HCO$^+$ \jtra10. The dynamical age of the outflow is $\le$
10,000 yr. The velocity gradient seen in H$^{13}$CN  is consistent with
Keplerian rotation.  Assuming that the gas is gravitationally bound, the
mass of the central object is $\sim$ 40 \Msun.  The mass of the
continuum ``disk'' is $\geq$ 100 \Msun\ and has a luminosity of 
$\sim$10$^4$ \Lsun. H$^{13}$CN gives a mass $\sim$400 \Msun\ for the
rotating disk, and $\sim$1000 \Msun\ for the extended
(20\arcsec {}) envelope. Our observations confirm that this is an extremely 
massive protostar in its earliest stages.

\end{abstract}


\section{Introduction}

The formation of high mass stars is still poorly understood, although
one expects a high mass protostar to form an accretion disk and drive
an outflow similar to what one sees in low mass protostars. Recent
surveys show that outflows are very common in high-mass star forming
regions \citep[and references therein]{Beuther02} but that accretion disks
have been far more elusive, even though they will have to be larger and
have larger velocity gradients than accretion disks around low mass
stars, which show velocity gradients of only a few tenths of
km~s$^{-1}$, see e.g. \citet{Mannings97}.  Although there have been a
number of papers reporting rotating disks around high-mass stars, most
of these show marginal, barely resolved systems. The best case to date,
the young high-mass star IRAS 20126$+$4104 \citep{Zhang98,Cesaroni97},
has a bolometric luminosity of 1.3 10$^4$ \Lsun\ and drives a massive
molecular outflow. 

In this letter we present the first results from high spatial
resolution BIMA observations of NGC\,7538\,S.  \citet{Sandell03a}
showed that  NGC\,7538\,S is a cold, very massive elliptical dust
source ($\sim$400\Msun) associated with H$_2$O and OH maser emission
but only weak free-free emission, which has all the characteristics of
a high-mass Class 0 source. The protostellar source is located in the
molecular cloud south-east of the large \ion{H}{2} region NGC\,7538,
which is known to contain several centers of active and on-going
high-mass star formation \citep{Werner79}. The distance to NGC\,7538 is
assumed to be 2.8 kpc.  NGC\,7538\,S is about 80\arcsec\ to the south
of the well studied ultracompact \ion{H}{2} region IRS\,1.

NGC\,7538\,S is more isolated than IRS\,1 and therefore easier to
study.  We confirm that NGC\,7538\,S is a high mass protostar and
resolve the disk around the protostar both in continuum and in
molecular lines. In this letter we present data that suggest that
the protostar is surrounded by a massive, rotating accretion disk,
which drives a very young and powerful outflow perpendicular to the
disk.

\section{Observations and Data Reduction}

The observations of NGC\,7538\,S were made with the BIMA array in
2001/2002 using two frequency settings in the B and C-array
configuration. The correlator was split into four 25 MHz bands
resulting in a velocity resolution of  $\sim$0.34 km~s$^{-1}$.  In
this letter we base most of our discussion on the frequency setting
that included H$^{13}$CN \jtra10, and HCO$^+$ \jtra10, which was
observed in both B- and C-arrays. The other frequency setting included
the molecular transitions HN$^{13}$C \jtra10, SiO  v=0 \jtra21 and
H$^{13}$CO$^+$ \jtra10. Only one C-array track was obtained in this
setting. The images therefore have poorer spatial resolution and
sensitivity, but are consistent with the deeper images seen in
H$^{13}$CN and HCO$^+$.

The data were reduced and imaged in a standard way using MIRIAD
software \citep{Sault95}. Phase calibration was applied using
observations of the quasar 0102$+$584 at intervals of 30 minutes. The
phase calibrator was observed using an 800 MHz bandwidth for 3 min.
3C454.3 was observed for 10 minutes as a bandpass calibration.  The
flux density scale was checked from observations of Mars. The
uncertainty in the absolute amplitude scale $\sim$15 \%, but the
relative amplitude of spectral lines within the same receiver tuning is
within $\sim$5 \%.

The data were imaged with weighting inversely proportional to the
variance in order to obtain the best signal to noise ratio.  Averaging
4 spectral channels we obtained an RMS noise level of 24 mJy (0.3 K)
with a synthesized beam FWHM 3\ptsec75 $\times$ 3\ptsec59 and peak
sidelobe level of -7 \%.  Spectral windows which did not contain any
significant spectral line emission were averaged to provide a continuum
image with an RMS $\sim$3 mJy. The continuum emission was subtracted
from the spectral line channels, and the images deconvolved using the
CLEAN algorithm.

\section{The continuum ``disk''}
\label{sect-cont}

The single dish sub-mm continuum maps \citep{Sandell03a} show
NGC\,7538\,S as an elliptical source embedded in a narrow dust filament.
With BIMA we filter out the extended cloud emission and detect only the
sub-mm source.  At 3.4~mm the continuum source is well resolved with a
peak on the northeastern side of an extended elliptical source (Fig 1). 
A two-component Gaussian fit shows that the northeastern peak  is an
unresolved point-like source with a flux density of 20 mJy offset 
3\ptsec7, 0\ptsec5 from the nominal sub-mm position. The position of the
3.4~mm point source coincides with the OH and H$_2$O maser and the 6~cm
VLA position \citep[O. Kameya 2003, private communication]{Argon00,Kameya90}. At
6~cm the VLA source has a flux density of 2.8~mJy. If the 6~cm emission
is due to an ionized wind, i.e. follows a $\nu^{0.6}$ frequency
dependence, about half of the 3.4~mm flux is an excess, presumably due
to hot dust close to the protostar. However, the free-free emission
could also have a steeper frequency dependence. In the following we
assume that the point-source emission is all due to free-free emission.

The extended elliptical source has a size of 13\ptsec7 $\times$
8\ptsec1 and a position angle (P.A) of 71\degr, offset $+$1\ptsec7, $-$0\ptsec4 from the
nominal sub-mm position with an integrated flux density of 190
mJy, i.e.  it agrees within errors both in size and position with the
sub-millimeter source. In this respect this source
differs from IRS\,1 and IRS\,9, which both appear unresolved at 3~mm
\citep{Tak00} and from nearby low mass protostars, which are only 
marginally resolved with a $\sim$0\ptsec5 beam \citep{Looney00}.

To estimate a mass from our continuum observations we adopt  a
single-temperature optically thin thermal dust model. The total mass of
gas and dust, M,  can then be expressed as  ${\rm M = S_\nu
D^2/(\kappa_\nu B_\nu(T_d))}$, where ${\rm B_\nu(T_d)}$ is the Planck
function, $\kappa_\nu$ is the dust mass opacity, $T_d$ is the dust
temperature, $D$ is the distance, and S$_\nu$ is the integrated flux
density at the frequency $\nu$.  We assume a gas to dust ratio of 100
and adopt a  dust mass opacity at 87 GHz, $\kappa_{87} $= 0.0073 $cm^2
g^{-1}$, corresponding to a dust emissivity, $\beta$ = 1. This value was
also used by \citet{Looney00} in their BIMA study of low mass
protostellar disks. With a dust temperature,T$_d$ = 35~K
\citet{Sandell03a}, we find the total mass of the continuum disk to be
100\Msun. We note, however, that the isothermal greybody fits by
\citet{Sandell03a} predicted a steeper emissivity law, $\beta$ = 1.6,
which would result in a five times smaller mass opacity and hence a more
massive disk.

\section{A rotating protostellar disk ?}
\label{sect-moldisk}

The map of integrated H$^{13}$CN \jtra10 emission (Fig. 1) shows a
bright elongated source surrounded by a more extended envelope
superposed on the 3.4~mm continuum. The elongated source in the center
of the map has  the same orientation and extent as the source we
see in dust continuum. For a two-component Gaussian fit, we find the
bright emission offset 2\ptsec0, 1\ptsec0 from the nominal sub-mm
position, i.e. within errors coincident with the 6~cm VLA and the OH/H$_2$O
maser position at +3\ptsec7, +1\ptsec0. The fitted size:  11\ptsec1
$\times$ 6\ptsec8 with a P.A. = 55 $\pm$ 5\degr,  is slightly smaller
than the  continuum  disk. The position velocity plot taken
along the major axis of the bright  H$^{13}$CN emission shows a clear
velocity gradient across the source (Fig 2). The emission is
red-shifted to the southwest and blue-shifted to the northeast with the
center of symmetry coinciding within 2\arcsec\ with the position of the
protostar. A rotation curve derived from these data is slightly
asymmetric with the central velocity close to $-$56 km s$^{-1}$. It
reaches a peak of $\pm$1.35 km~s$^{-1}$ at a radius of 5\arcsec\
(14000 au) in both directions. 
The observed velocity gradient corresponds to an enclosed 
mass of $\sim$ 30 / ${\rm sin}^2i$ \Msun\ for a rotating disk
to be gravitationally bound. With an assumed inclination, i =
60\degr\ (see below) the mass is $\sim$ 40
\Msun.

H$^{13}$CO$^+$  \jtra10 and  HN$^{13}$C \jtra10  were observed 
with only one C-array track and therefore lack the spatial resolution to
resolve the disk. Both molecules show strong extended emission towards
the protostar and we can therefore use all three molecules to get an
estimate of the mass of the disk and the surrounding molecular
envelope.  For estimating column densities and masses we assume LTE and
that the excitation temperature for all molecules is the same as that
for the dust, i.e. 35~K. The permanent dipole moments are taken from
\citet{Blake87}, except for HCO$^+$, for which we adopt $\mu$ = 4.07
Debye \citep{Haese79}. For total mass estimates we use abundance
ratios to  H$_2$  similar to those in the OMC-1 extended ridge
\citep{Blake87}, i.e. [HCO$^+$]/[H$_2$] = 2 10$^{-9}$, [HCN]/[H$_2$] =
5 10$^{-9}$, and [HNC]/[H$_2$] = 5 10$^{-10}$, and the isotope
ratio, [$^{12}$C]/[$^{13}$C] = 85 \citep{Wilson94}. For H$^{13}$CN we
get a peak column density of 7 10$^{13}$ cm$^{-2}$ for the disk and
5 10$^{13}$ cm$^{-2}$ for the surrounding envelope, resulting in a
disk mass $\sim$ 400 \Msun\ and an envelope $\sim$ 1000 \Msun. For
H$^{13}$CO$^+$ and HN$^{13}$C we do not have enough spatial resolution
to separate the disk and the envelope, but we derive combined masses
$\sim$ 1000 and 3000\Msun, for H$^{13}$CO$^+$ and HN$^{13}$C,
respectively.

\section{The high velocity outflow }
\label{sect-flow}

We see clear high velocity wings in both HCO$^+$ \jtra10 and SiO v=0
\jtra21. Towards the protostar the HCO$^+$ shows dominantly
blue-shifted high velocity emission extending 29 km s$^{-1}$ from the
systemic velocity ($\sim$ $-$56 km $s^{-1}$) with a sharp, deep
absorption on the red-side at nearby cloud velocities, suggesting
infall of the surrounding cooler envelope. The SiO line wings at the
position of the disk appear more symmetric, presumably
due to the broader beam and the absence of self-absorption. Both SiO
and HCO$^+$ show the same distribution of the high velocity gas. Here
we only present the results from our HCO$^+$ imaging, which has better
signal-to-noise and higher spatial resolution.

HCO$^+$ shows a strong, compact, jet-like bipolar high velocity outflow
at a pa of 147\degr $\pm$ 5\degr\ (Fig 3). The symmetry axis of the
outflow passes through the VLA source. The H$^{13}$CN map shows that
the surrounding cloud core is more dense to the north than to the
south. This is also evident from the outflow, which is much more
compact to the northwest than to the southeast. In the northwest the
outflow is blue-shifted and terminates sharply at $\sim$7\arcsec\ from
the protostar. To the southeast the outflow is mostly red-shifted and
well collimated  (aspect ratio $\sim$ 2:1) to $\sim$16\arcsec\ from the
protostar, where it expands to become a more wide angle flow,
presumably breaking out of the cloud at $\sim$ 30\arcsec\ from the
protostar. The strongest blue-shifted emission is seen at the tip of
the northwestern outflow lobe with outflow velocities extending to 27
km s$^{-1}$ from the systemic cloud velocity.  The southeastern outflow
lobe shows both red- and blue-shifted emission with outflow velocities of
$\sim$24 km~s$^{-1}$ in the red and blue-shifted emission up to 18
km~s$^{-1}$ in the dense, well collimated part of the outflow. At the
tip of the outflow, the outflow velocities are lower, presumably
because the high velocity gas is too tenuous to be excited in HCO$^+$.

Since there is both blue- and red-shifted emission to the southeast ,
and little overlap between the outflow lobes at the star, the outflow
must have fairly high inclination.  In the following we assume an
inclination of 60\degr, which is roughly consistent with the observed
aspect ratio of the disk, i.e. we see the disk almost edge on. The
northwestern blue-shifted outflow lobe is still confined by the
surrounding cloud, while the southeast outflow appears to have broken
through the dense cloud core.  Since the extent of the outflow to the
southeast is less well defined we derive outflow characteristics mainly
from the northwest blue-shifted outflow lobe.

The blue-shifted outflow lobe gives a dynamical time scale of
$\sim$2,000 yr, while the red-shifted emission gives $\sim$10,000 yr,
corrected for a 60\degr\ inclination. Note that the velocities are
probably much higher in the red-shifted outflow, since it is expanding
through a less dense cloud. We assume  HCO$^+$ to be optically thin
with an excitation temperature of 30~K, the same as \citet{Shepherd96}
used in their study of outflows from high mass stars.  If we further
assume a normal HCO$^+$ abundance in the outflow (Section
~\ref{sect-moldisk}),  we find a mass of 9 \Msun\ for the
northwest blue-shifted outflow and 8 \Msun\ for the redshifted gas
in the southeast outflow lobe. The mass weighted momentum flux, $
{\rm F = mv/t_{dyn}}$ and the mechanical luminosity, $L_{mec} =
mv^3/2r$, where r is the extent of the outflow, is 90  10$^{-3}$ {\rm
$M_{\odot} km s^{-1} yr^{-1}$, and 130 \Lsun\ for the blue outflow
alone. The mass loss rate, based only on the confined blue-shifted
outflow, could be as high as $\sim$5~10$^{-3}$ {\rm
$M_{\odot}~yr^{-1}$.

\section{Discussion and Conclusions}
\label{sect-discussion}

The new data we present in our letter reveal a well resolved elliptical
source mapped in several optically thin lines as well as in 3.4 mm
continuum. The peak of the line emission is centered within 1\arcsec\ -
2\arcsec\ of the VLA 6cm continuum, OH, and H$_2$O maser emission
position.  The 3.4 mm continuum emission has an extent similar to the
elongated source seen in optically thin high-density tracers with a
peak (hotspot) within 1\arcsec\ of the VLA and maser position.
Follow-up observations at 1~mm in both line and continuum
(Sandell, Wright, \& Forster 2003, in preparation) show the disk more clearly and show that CH$_3$CN
\jtra{12}{11}, which traces hot gas, is centered within 0\ptsec2 of
this position. FIR observations \citep{Werner79,Thronson79} show a FIR
source with a luminosity of $\sim$ 10$^4$ \Lsun, with a narrow FIR
spectral energy distribution, strongly suggesting that the FIR
luminosity is produced by a single central source.  The FIR source has
no near or mid-IR counterpart suggesting that it is heavily obscured.
The FIR observations do not have enough spatial resolution or
positional accuracy to confirm that the luminosity originates from the
the OH/H$_2$O masers, but all 1665 MHz OH masers require a luminous
source for their excitation; therefore we conclude that the source
powering the OH maser is the origin for the bulk of the observed FIR
luminosity.

The H$^{13}$CN map (Fig 2) shows a clear velocity gradient along the
major axis of the elliptical source. If we interpret this as a rotating
disk gravitationally bound by the central source, we find that the
center of symmetry is within $\le$ 2\arcsec\ of the maser source.
Since this position also lies on the symmetry axis of the outflow,
defined by our high spatial resolution HCO$^+$ map, there is no doubt
that this is the center of the mass and the protostar. From the
observed velocity gradient we derive an enclosed mass of 40 \Msun
(Section ~\ref{sect-moldisk}). Since we have resolved the emission of
the disk from the envelope, we can use the integrated emission of the
disk to derive a mass. Assuming normal abundances and a kinetic
temperature of 35~K for the disk we find a mass $\sim$ 400 \Msun from
the H$^{13}$CN, and $\sim$ 100 \Msun from the 3.4~mm continuum
emission.  Considering the uncertainties in molecular abundances and
dust opacity, these estimates are in reasonable agreement. The mass of
the central object and the disk are of the same magnitude.

The bipolar HCO+ high velocity outflow (Section ~\ref{sect-flow}) may be
driven by the rotating protostellar disk.  The outflow is extremely
young; the dynamical timescale from the blue-shifted outflow indicates
an age of $\leq$ 2000 yr, with an upper limit of 10,000 yr derived from
the red-shifted outflow. \citet{Shepherd96} show that there is a clear
correlation between mass loss rate and bolometric luminosity of the
central source over a wide range of luminosities. If we use the
correlation derived by \citet{Shepherd96}, our observed mass loss rate
corresponds to a luminosity of close to 10$^5 $\Lsun, while the
observed luminosity is only 10$^4 $\Lsun. Our observations therefore
suggest an elevated mass loss rate in the very earliest stages of the
formation of a massive star. Since we find the rotating accretion disk
to be more massive than the central protostar, the disk will be highly
unstable and is likely to feed both the outflow and the central
accreting protostar. Model calculations by \citet{Yorke95} for a
somewhat less massive central star (10 \Msun{}), find that the
accretion disk evolves into a thin flared disk with a ratio of
M$_\star$/M$_{disk}$ $\sim$ 4, for all the models that they computed.
Even though they ignored magnetic fields, these models should still
give an idea of how the disk is expected to evolve.

There is only one near-IR source in the vicinity of NGC\,7538\,S, IRS\,11
(Fig 3). This IR source is associated with nebulosity and a knot of
vibrationally excited H$_2$ emission, suggesting that it is most likely
a young star. There are several additional H$_2$ knots in the vicinity
of NGC\,7538\,S;  two appear to be associated with the molecular
outflow. The only other evidence for young objects in this region is 
an unexplained high velocity HCO$^+$ emission feature 10\arcsec\ southwest
of the protostar, three vibrationally excited H$_2$ knots $\sim$ 20\arcsec\
east of it, and an H$_2$O maser $\sim$ 30\arcsec\ west of it. The high
velocity HCO$^+$ could be part of a wide angle disk-wind or due to a
young low luminosity source in the outskirt of the disk. This could
explain the observed asymmetry in the continuum disk (more extended to
the southwest), but it cannot produce the observed velocity gradient. 
The eastern H$_2$ knots and the western H$_2$O maser are unrelated to
the protostellar source.  

It is therefore possible that the extended disk/envelope system
harbours yet another protostar. Such a protostar would have much lower
luminosity and mass, since the massive protostar can explain most, if
not all of the observed luminosity. Although high-mass stars usually
form in clusters, all the evidence we have so far, suggest that
NGC\,7538\,S is an isolated high mass star. If the central protostar is
a single star, it may evolve into a late O or early B-star.

\acknowledgements 

We thank Dr. Osamu Kameya for sharing his unpublished VLA results with
us. The BIMA array is operated by the Universities of California
(Berkeley), Illinois, and Maryland with support from the National
Science Foundation.}

{}

\plotone{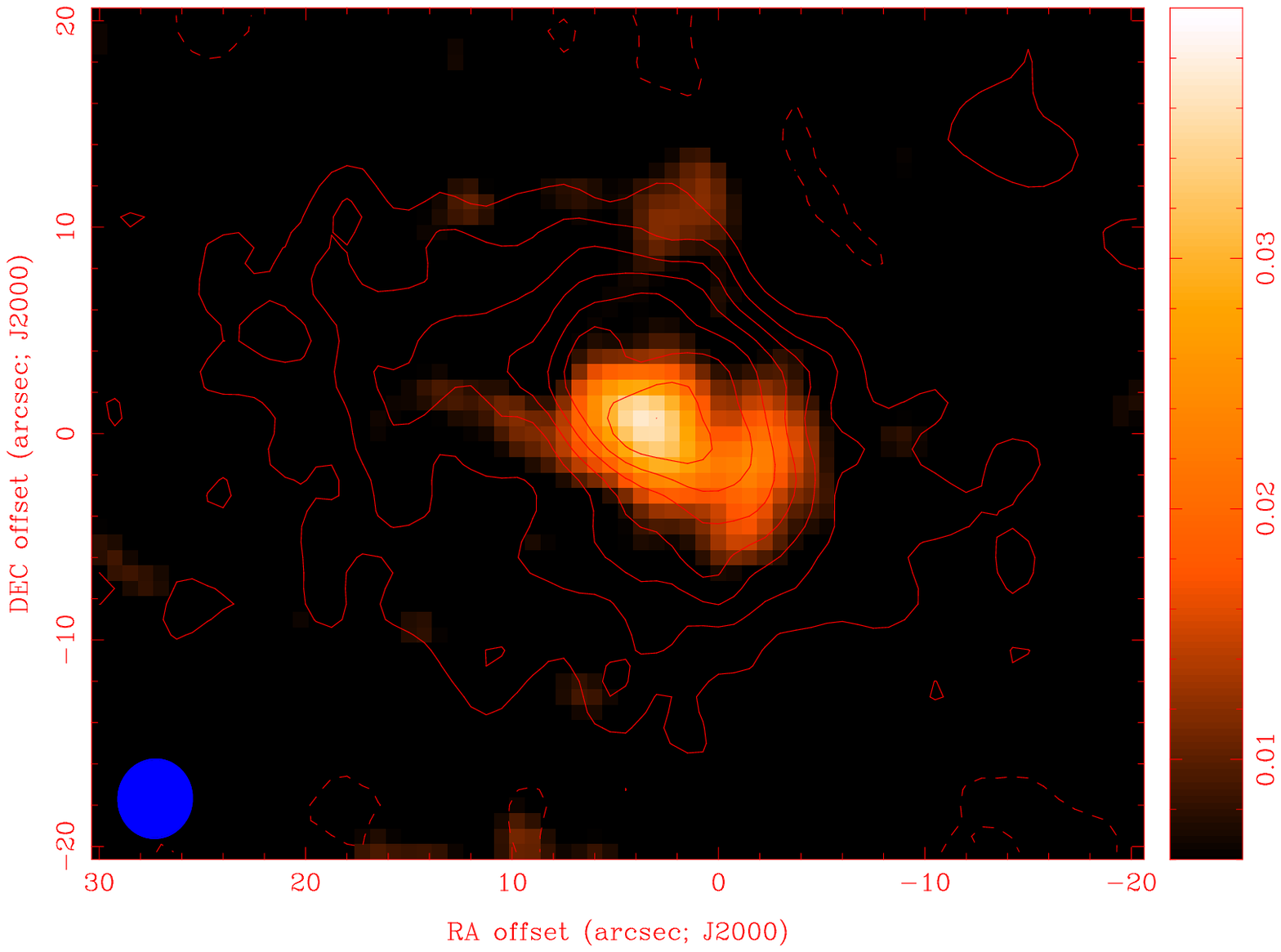}
\figcaption[]{Color image of 3.4~mm continuum emission of NGC\,7538\,S
overlaid with contours of H$^{13}$CN \jtra10 emission integrated over
the velocity range $-$70 to $-$40 km~s$^{-1}$. The maps are offset
($-$3\ptsec7,$-$0\ptsec5) from the 3.4~mm peak: $\alpha$(2000.0) =
23$^h$  13$^m$ 44\psec98, $\delta$(2000.0) = $+$ 61\degr\
26\arcmin\ 49\ptsec2. The intensity scale for the continuum is in
Jy~beam$^{-1}$. The contours for the H$^{13}$CN emission go from $-$6 K km
s$^{-1}$ (dotted contours) and thereafter from 6 K km s$^{-1}$ with
steps of 6 K km s$^{-1}$. The beam FWHM is plotted in blue.}

\plotone{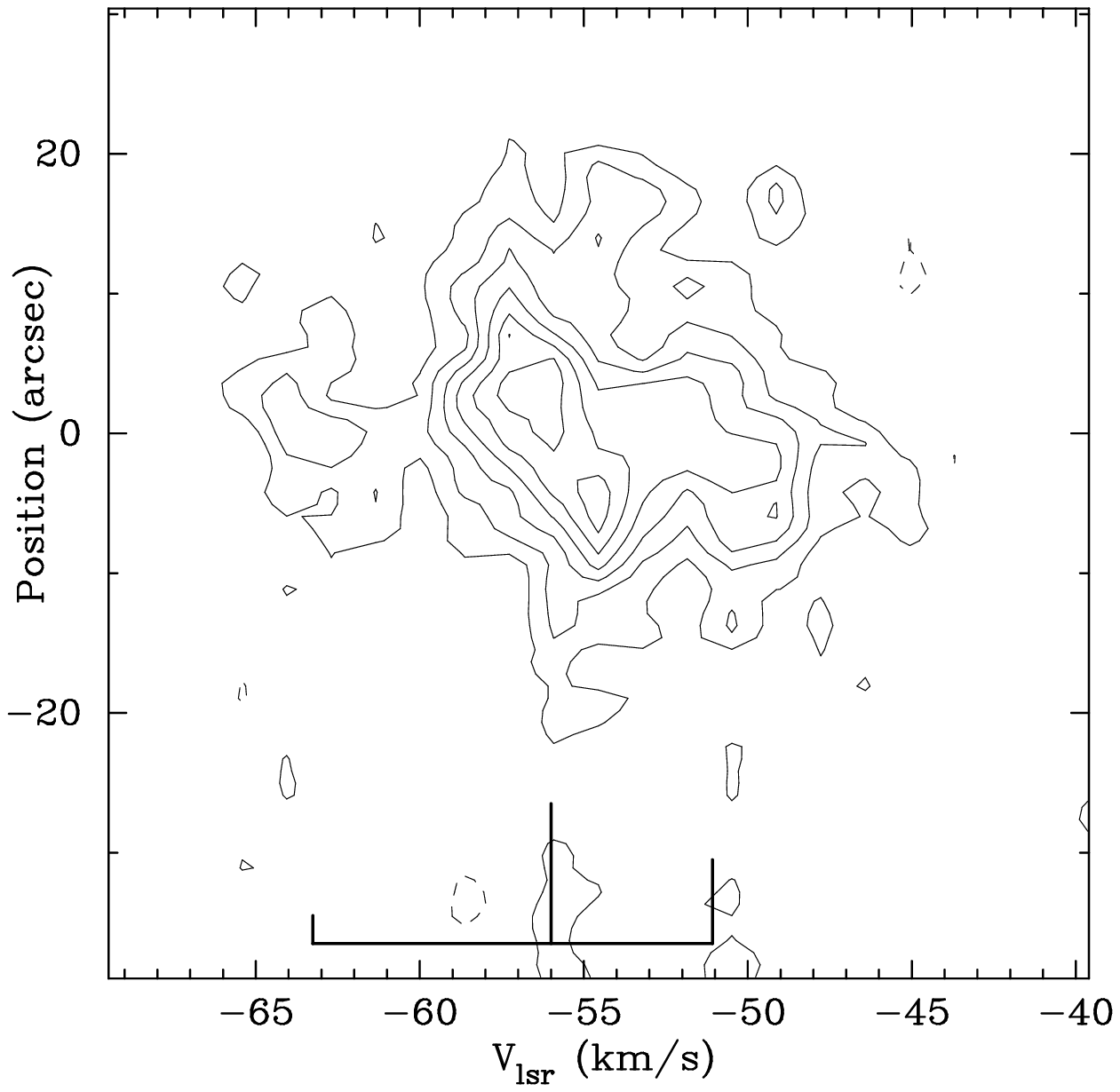}
\figcaption[]{Position velocity plot of the H$^{13}$CN \jtra10 emission
along the disk plane at pa = 60.0 \degr. Positive offsets are to the
northeast. The velocity gradient is seen in all three hyperfine lines.
The velocity scale is relative to the strongest hyperfine component, F
= 2 - 1, centered at a systemic velocity of $\sim$ $-$56 km s$^{-1}$.
The line drawing at the bottom of the plot shows the position of three
hyperfine components and their relative intensity. The disk emission is
red-shifted to the southwest  and blue-shifted to the northeast (see
text).}

\plotone{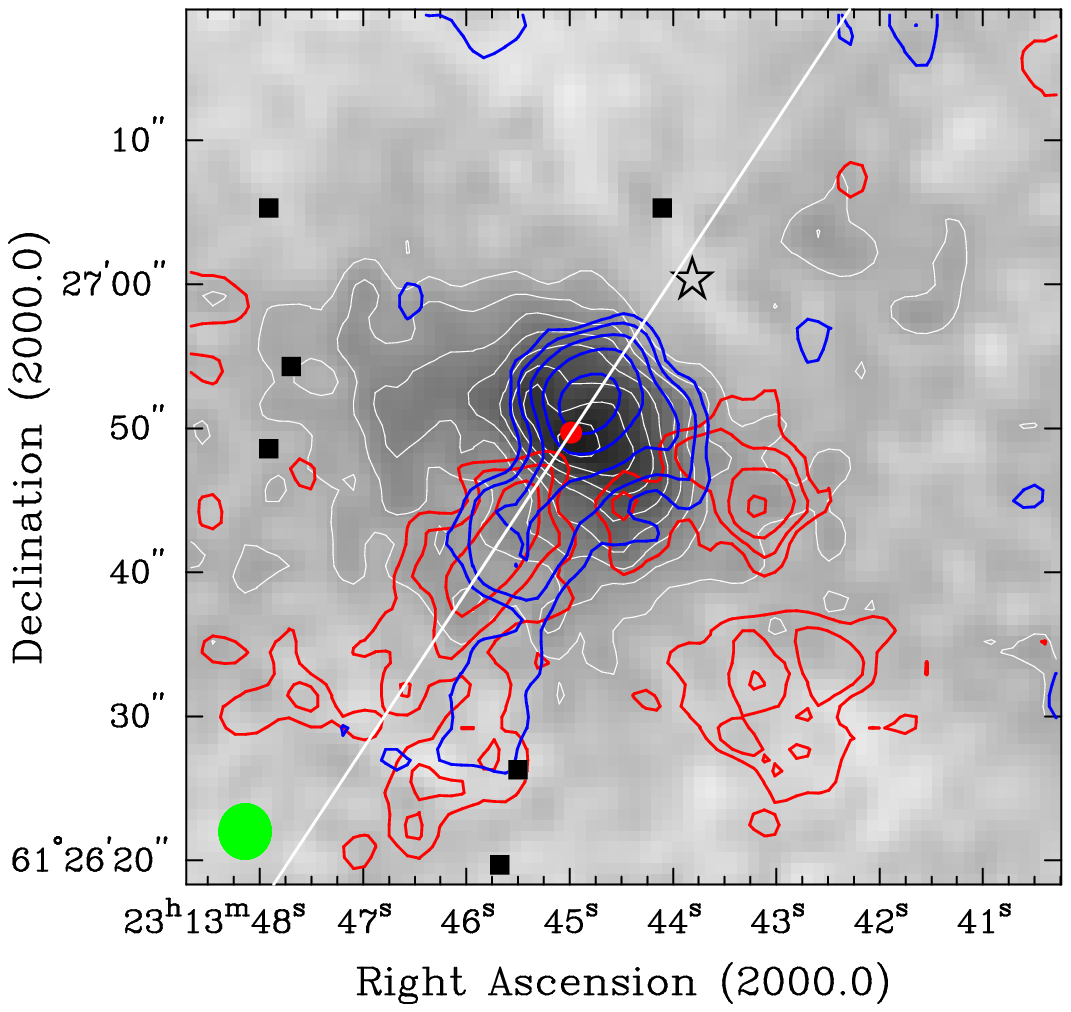}
\figcaption[]{Contour map of blue- and red-shifted high velocity
HCO$^+$ \jtra10 integrated over 20 km s$^{-1}$ overlaid on a greyscale
image of  integrated H$^{13}$CN. The red circle marks the position of
the protostar and the white line shows the symmetry axis of the
outflow. The star symbol marks the position of IRS\,11 and the filled
squares are H$_2$ knots from \citet{Davis98}. The contours are
logarithmic and start at 4.6 K km s$^{-1}$ for HCO$^+$ with a step of
10$^{0.2N}$, where N= 1,2.. The beam FWHM is plotted in green.}

\end{document}